\documentclass[conference]{IEEEtran}
\IEEEoverridecommandlockouts
\usepackage{cite}
\usepackage{amsmath,amssymb,amsfonts}
\usepackage{algorithmic}
\usepackage{graphicx}
\usepackage{textcomp}
\usepackage{xcolor}
\usepackage{subcaption}
\usepackage{booktabs}
\usepackage{array}
\usepackage{url}
\usepackage{hyperref}

\def\BibTeX{{\rm B\kern-.05em{\sc i\kern-.025em b}\kern-.08em
    T\kern-.1667em\lower.7ex\hbox{E}\kern-.125emX}}

\usepackage{float}

\begin{document}

\title{Unsupervised Rhythm and Voice Conversion of Dysarthric to Healthy Speech for ASR
\thanks{This work was partially supported by the Swiss National Science Foundation grant agreement no. 219726 on ``Pathological Speech Synthesis (PaSS)'' and the Innosuisse flagship grant agreement no. PFFS-21-47 on ``Inclusive Information and Communication
Technologies (IICT)''.}
}

\author{\IEEEauthorblockN{Karl El Hajal\IEEEauthorrefmark{2}\IEEEauthorrefmark{3},
Enno Hermann\IEEEauthorrefmark{2}, Ajinkya Kulkarni\IEEEauthorrefmark{2} and
Mathew Magimai.-Doss\IEEEauthorrefmark{2}}
\\
\IEEEauthorblockA{
\IEEEauthorrefmark{2}Idiap Research Institute, CH-1920 Martigny,
Switzerland\\
\IEEEauthorrefmark{3}EPFL, \'Ecole polytechnique f\'ed\'erale de Lausanne, CH-1015 Lausanne, Switzerland}}

\maketitle

\begin{abstract}
Automatic speech recognition (ASR) systems are well known to perform poorly on dysarthric speech. Previous works have addressed this by speaking rate modification to reduce the mismatch with typical speech. Unfortunately, these approaches rely on transcribed speech data to estimate speaking rates and phoneme durations, which might not be available for unseen speakers. Therefore, we combine unsupervised rhythm and voice conversion methods based on self-supervised speech representations to map dysarthric to typical speech. We evaluate the outputs with a large ASR model pre-trained on healthy speech without further fine-tuning and find that the proposed rhythm conversion especially improves performance for speakers of the Torgo corpus with more severe cases of dysarthria. Code and audio samples are available at \url{https://idiap.github.io/RnV}.
\end{abstract}


\section{Introduction}
Speech disorders can significantly impact a person's ability to communicate effectively. Dysarthria, a common speech disorder, is caused by neurological impairments that weaken or disrupt the muscle control required for speaking. This results in challenges in articulation, rhythm, pitch, and volume compared to healthy speakers~\cite{Duffy2012}. As a result, speech technologies such as automatic speech recognition (ASR) systems, which are generally trained on typical healthy speech, struggle to accurately process dysarthric speech~\cite{Moore2018}.
There is considerable interest in developing assistive technologies tailored for speakers with dysarthria. However, creating these systems is challenging. Dysarthric
speech varies significantly from one speaker to another, and the availability of
speech data is limited, as data collection can be exhausting for individuals
with this condition. To address this limitation, researchers have explored
data generation methods such as text-to-speech (TTS)
for synthesizing dysarthric speech~\cite{Soleymanpour2022,Hermann2023,Leung2024} and voice conversion (VC) to transform healthy
speech into dysarthric speech, or vice versa.

Lower speaking
rates especially can degrade ASR performance because acoustic model
architectures are optimized for typical speech. Previous works
address this, for example, by adjusting frame shift during feature
extraction~\cite{Espana-Bonet2016}, normalizing speaking rates at train or test
time~\cite{Bhat2016,Xiong2019}, or augmenting with rate-adjusted or converted
typical speech~\cite{Vachhani2018,Halpern2021,Prananta2022}. A common downside to these
approaches is that the global or phoneme class-specific speaking rate factors
are estimated based on forced alignments, which require transcribed speech data
that may not be available for new speakers at test time.

Recent studies show that high-quality zero-shot voice
conversion~\cite{baas23_interspeech} and morphing~\cite{hajal2024ssltts} can be
achieved through unsupervised methods by leveraging pre-trained self-supervised
learning (SSL) speech representations. Indeed, discrete speech features from
speech SSL encoders offer a useful property: units from different speakers with
the same phonetic content tend to cluster closely~\cite{baas23_interspeech}.
This allows mapping units from a source speaker to the closest matching units of
a target speaker, enabling effective voice conversion, including for stuttered speech~\cite{knnvc_followup}.

A primary advantage of these approaches is their minimal data requirements. Since the conversion is zero-shot and unsupervised, no large datasets or fine-tuning are needed. This low-data requirement is particularly well-suited for the dysarthric speech domain. However, a limitation of this technique is its inability to capture the target speaker's rhythm, as conversion is achieved through frame-by-frame replacement which keeps the original rhythm unchanged.
In parallel, an unsupervised method to model the rhythm characteristics of different speakers has been proposed~\cite{urhythmic}. This approach uses soft-speech units generated by a soft content encoder, which predicts a distribution over discrete units from a self-supervised learning (SSL) speech encoder. A clustering algorithm categorizes audio segments into three speech types and models the duration of these types for different speakers. Rhythm can then be transferred from a source speaker to a target speaker by mapping the source distribution for each speech type to the target distribution and time-stretching the corresponding segments.

\begin{figure*}[h]
    \centering
    \includegraphics[width = .65\textwidth]{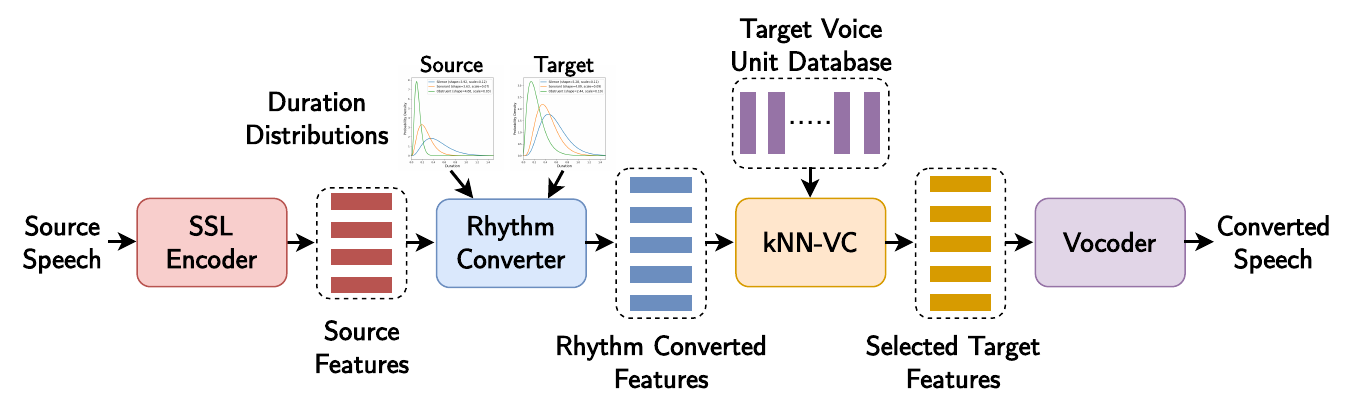}
    \caption{Overview of the unsupervised Rhythm and Voice conversion framework.}
    \label{fig:model_architecture}
\end{figure*}

This work investigates whether such unsupervised methods can  model the rhythmic characteristics of both healthy and dysarthric speech, and consequently convert dysarthric speech from the Torgo corpus~\cite{Rudzicz2012_torgo} into typical speech.

We introduce a Rhythm and Voice (RnV) conversion framework, which leverages self-supervised embeddings to model and convert rhythm and voice characteristics in an unsupervised manner. By combining and adapting the aforementioned methods, we explore their application in the dysarthric speech domain. We convert dysarthric speech of varying intelligibility into healthy speech and assess the impact of rhythm modeling and voice conversion on the resulting speech in terms of ASR performance with Whisper~\cite{whisper}, a large ASR model pre-trained on typical speech.

\section{Methods}
The Rhythm and Voice (RnV) conversion framework is visualized in Figure \ref{fig:model_architecture}.
Within this framework, an SSL speech encoder first extracts features from a source utterance and a set of utterances from a target speaker. Clustering techniques are applied to the target utterances to model the target speaker's rhythm characteristics, facilitating rhythmic conversion by mapping the source utterance to the target’s duration characteristics.
Additionally, a kNN-VC model~\cite{baas23_interspeech} converts the voice by comparing individual frames from the input utterance to frames from the target speaker's utterances, selecting the closest matches.
Both conversion methods rely on the same unsupervised speech representations.
Finally, a pre-trained vocoder decodes the converted speech representations back into waveforms.

\textbf{Speech Representations:} The speech representations required for this framework should support the unsupervised voice and rhythm conversion approaches we aim to use. Therefore, we need a representation that encodes utterances into discrete units, ensuring that frames with similar phonetic content from different speakers are close in linear space. Additionally, speaker information must be preserved in these features to enable accurate waveform reconstruction by a general pre-trained vocoder.

\textbf{Rhythm Modeling:} Recent works in dysarthric speech detection and analysis underscore the effectiveness of syllabic segmentation of dysarthric speech and its analysis using syllable-level features \cite{hovsepyan_syllabel_features}. This aligns with neurophysiologically plausible processes underlying human speech perception, where the human auditory system first decomposes sound into frequency components in the cochlea, then segments auditory information into syllabic units for cortical processing.

In the context of rhythm modeling, syllables also serve as natural units that reflect the timing and flow of speech. Indeed, by analyzing the pronunciation duration of individual syllables and syllable groups we can model speaker-specific rhythm and tempo \cite{pfau_speaking_rate}. However, segmenting speech into accurate syllabic segments typically requires time-aligned transcriptions, which are often unavailable in a VC setting.
Therefore, we adapt Urhythmic~\cite{urhythmic}, an unsupervised method for rhythm modeling, which leverages self-supervised representations to segment speech such that syllable rate and segments can be approximated. For compatibility with kNN-VC, we modify the method to use discrete speech units that contain speaker information rather than discard it, and a general pre-trained vocoder rather than a vocoder fine-tuned to each speaker. This enables any-to-any conversion rather than any-to-one.

The first step for Urhythmic is to obtain a segmenter by clustering discrete units from a speaker's utterances into 100 centroids using KMeans. Then, hierarchical clustering groups these centroids into three primary speech types. The speech type with the biggest overlap with silence sections, as determined by a voice activity detector, is classified as Silences, the type with the biggest overlap with voiced sections is determined to be Sonorants, and the third corresponds to Obstruents.
After performing this step, any given input utterance can be segmented by first calculating the distance of each input frame to each of the 100 centroids, and determining the log probability for each class. Using the dynamic programming algorithm presented in \cite{urhythmic}, consecutive frames are merged into longer segments, each representing one of the three main classes. This optimization process aims to form the longest possible segments, with a penalty parameter $\gamma$ used to encourage longer segments.
We consider two approaches to rhythm modeling: global and fine-grained.

\textit{Global:} This approach estimates each speaker's speaking rate, and converts a source speaker utterance's rhythm to a target rhythm by comparing speaking rates and time stretching the entire utterance accordingly in the discrete unit space. Speaking rate, which is typically calculated in terms of syllables-per-second, can be approximated by counting sonorants-per-second (since sonorants or vowels correspond to syllable nuclei), which was indeed shown to correlate well with the ground truth speaking rate \cite{pfau_speaking_rate, urhythmic}.

\textit{Fine-grained:} To model rhythm in more fine-grained fashion, we segment each speaker's utterances, tally the segment durations for each speech type, and fit a gamma distribution to these durations. This provides a duration distribution for each speech type per speaker. To convert one speaker's rhythm to another's, we segment the input utterance and adjust each segment’s duration to match the target’s distribution. This is done by calculating the Cumulative Distribution Function (CDF) of the source duration and applying the Percent Point Function (PPF) to determine the corresponding target duration with the same probability rank. This method preserves the source duration's rank within the target distribution. Finally, each source segment is time-stretched using linear interpolation to match its target duration.

\textbf{Voice conversion} is achieved through kNN-VC \cite{baas23_interspeech}, where each frame of the source speaker’s utterance is compared to frames in the target speaker's discrete unit database using cosine distance. The $k$ closest matches are selected and averaged to form a new converted unit, replacing the source frame. This method has proven effective in zero-shot scenarios, achieving high speaker similarity while preserving intelligibility. However, since conversion is performed through frame-by-frame replacement, the speaking rate is fixed by the source utterance and does not change for different target speakers \cite{hajal2024ssltts}. As a result, rhythmic characteristics specific to different speakers, which are critical in the dysarthric domain, are not modeled, hence the importance of rhythm modeling as a first step.

\vspace{-4pt}
\section{Experimental Setup}

We convert dysarthric speech into typical speech, tailoring it for ASR systems pre-trained on healthy speech without further fine-tuning. This section details our implementation of the conversion methods, the datasets used, the rhythmic analysis conducted, and the ASR configuration and evaluation.

\vspace{-3pt}
\subsection{Framework implementation}
We encode speech utterances using the 6th layer of WavLM-Large \cite{wavlm} as it effectively captures both speaker and linguistic phonetic information \cite{baas23_interspeech, hajal2024ssltts}. A pre-trained HiFi-GAN V1 vocoder \cite{hifigan} with a checkpoint from \cite{baas23_interspeech}, trained using their pre-matched paradigm, reconstructs the waveforms. In rhythm modeling segmentation, we set $\gamma = 3$ to promote longer segment calculations. For kNN-VC voice conversion, we retrieve the $k=8$ closest matches for each source unit and apply weighted averaging based on the cosine distance between each frame and the source frame.

\vspace{-3pt}
\subsection{Datasets}

We evaluate on the Torgo database \cite{Rudzicz2012_torgo}, which includes recordings from 7 control speakers and 8 speakers with dysarthria associated with Amyotrophic Lateral Sclerosis (ALS) or Cerebral Palsy (CP). Speakers with dysarthria are categorized by severity levels: severe, mod-severe, moderate, and mild. The dataset contains 2,340 single words and 725 sentences, ranging from 4 to 14 words, with diverse syntactic forms and syllabic structures. Speech was recorded using both head-mounted microphones and microphone arrays; we use the head-mounted microphone recordings. For the conversion target, we select the LJSpeech dataset~\cite{ljspeech}, which includes 24 hours of English audiobook recordings from a single speaker. All audio is resampled to 16kHz and normalized to a loudness of -20dB.

\subsection{Rhythm Analysis}

We first perform a rhythm modeling analysis, using our unsupervised method to segment dysarthric speech from Torgo, calculate each speaker's speaking rate, and determine per-speech type duration distributions. The results are visualized to assess the method's effectiveness in segmenting dysarthric speech and modeling its rhythmic characteristics.

\subsection{ASR Evaluation}

For ASR evaluation, we use the Whisper base model \cite{whisper}, a transformer-based encoder-decoder architecture pre-trained on 680,000~hours of healthy speech, encompassing various languages, accents, and acoustic conditions. We evaluate the impact of different conversion setups on ASR performance by calculating and comparing the overall word error rate (WER)\footnote{https://github.com/jitsi/jiwer} aggregated by dysarthria severity levels.
For conversion, we first train a segmenter on the LJSpeech dataset and compute both global and fine-grained rhythm models for LJSpeech. We then calculate rhythm models for each Torgo speaker. Next, we convert Torgo recordings to LJSpeech using different configurations and evaluate the impact of each configuration on ASR performance. We compare the following setups: original recordings (the baseline in this context), vocoded samples (original data encoded and resynthesized), rhythm-converted data using each rhythm model, voice-converted data using kNN-VC, and rhythm and voice converted data.

\begin{figure}[h]
    \centering
    \includegraphics[width = \columnwidth]{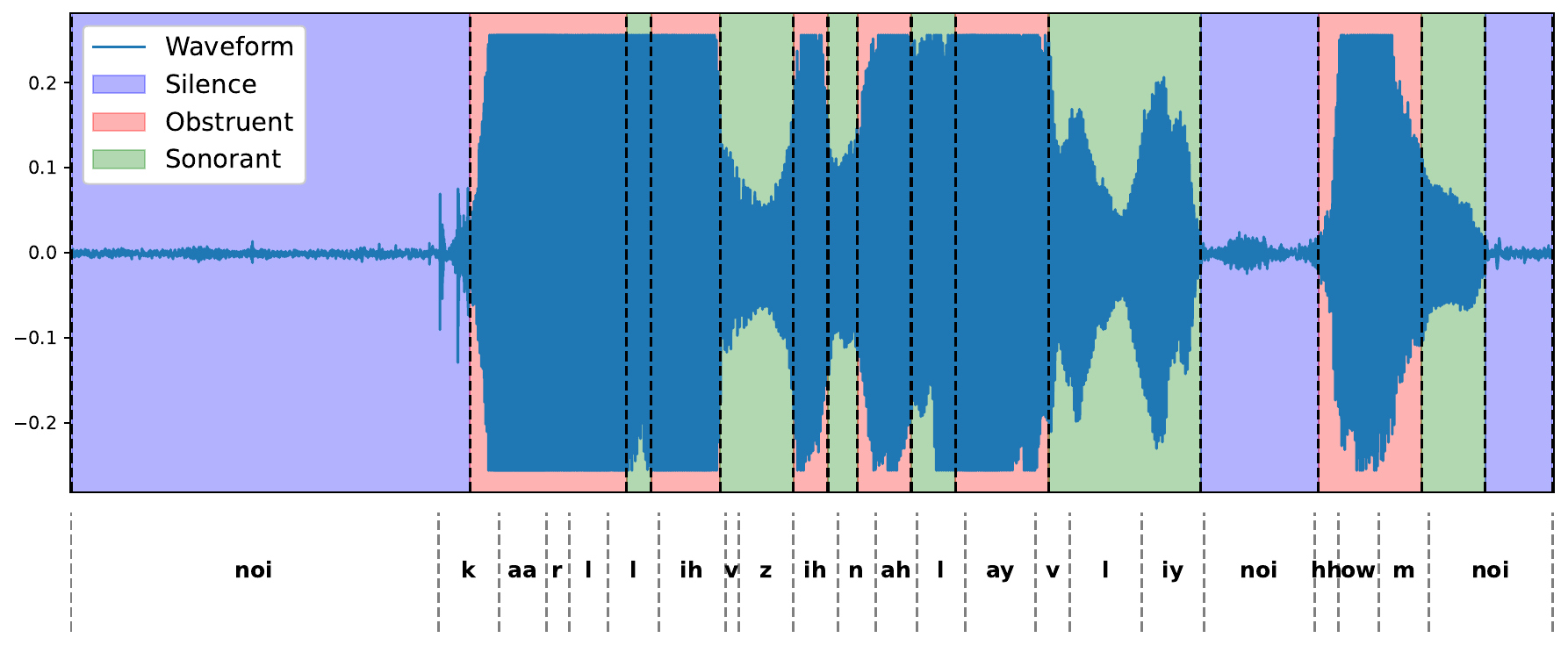}
    \caption{Segmented waveform of speaker M02 pronouncing the sentence “Carl lives in a lively home”. Ground truth phonemic transcriptions are shown at the bottom for reference ('noi' corresponds to noise).}
    \label{fig:segmented_waveform}
\end{figure}
\begin{figure*}[t]
  \centering
  \begin{subfigure}{.49\columnwidth}
    \includegraphics[width=\columnwidth]{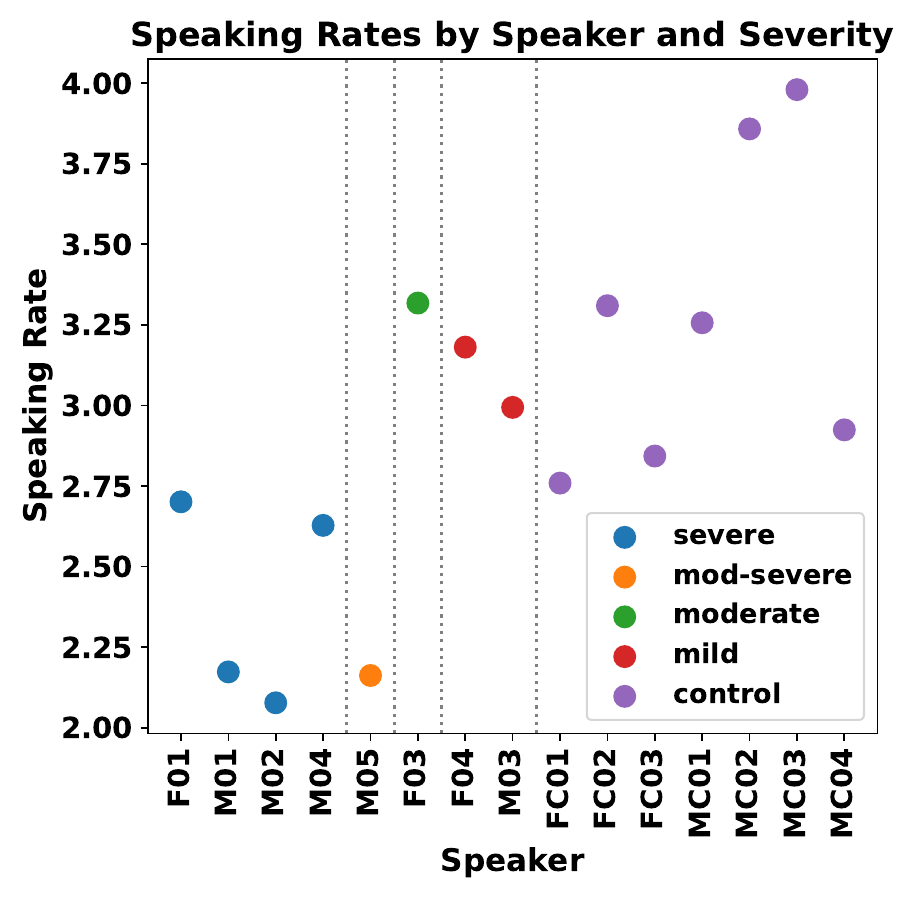}
    \caption{}
    \label{fig:speaking_rates}
  \end{subfigure}
  \begin{subfigure}{.49\columnwidth}
    \includegraphics[width=\columnwidth]{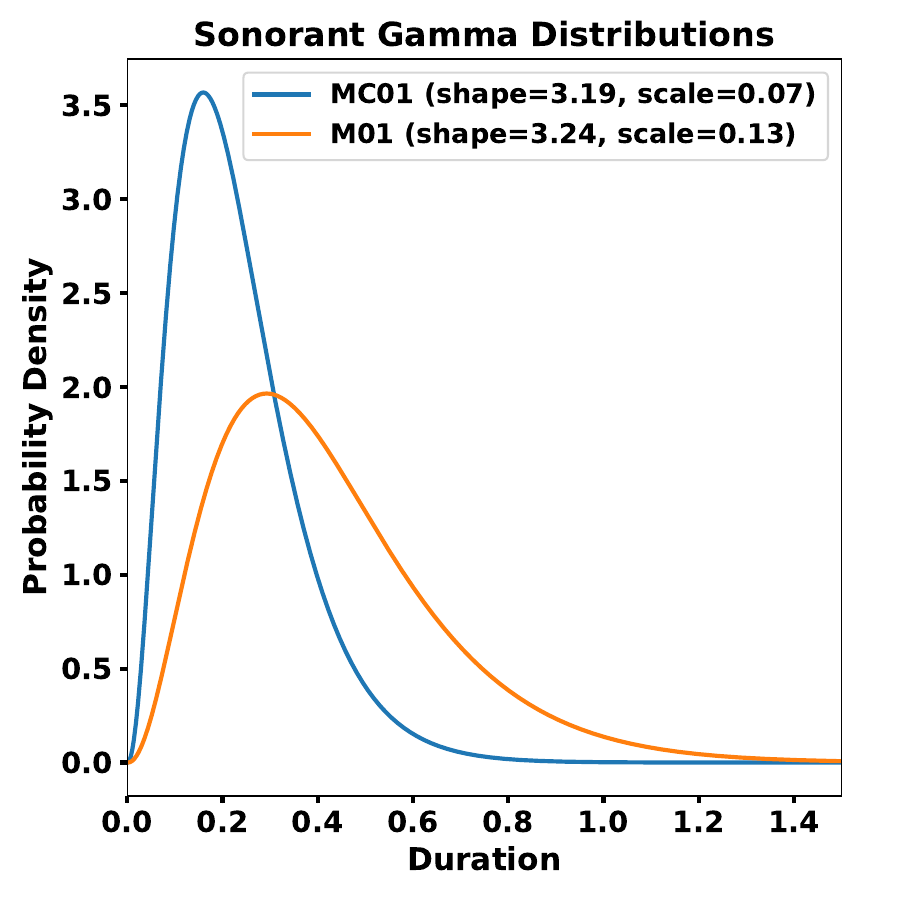}
    \caption{}
    \label{fig:sonorant_distribution}
  \end{subfigure}
   \begin{subfigure}{.49\columnwidth}
    \includegraphics[width=\columnwidth]{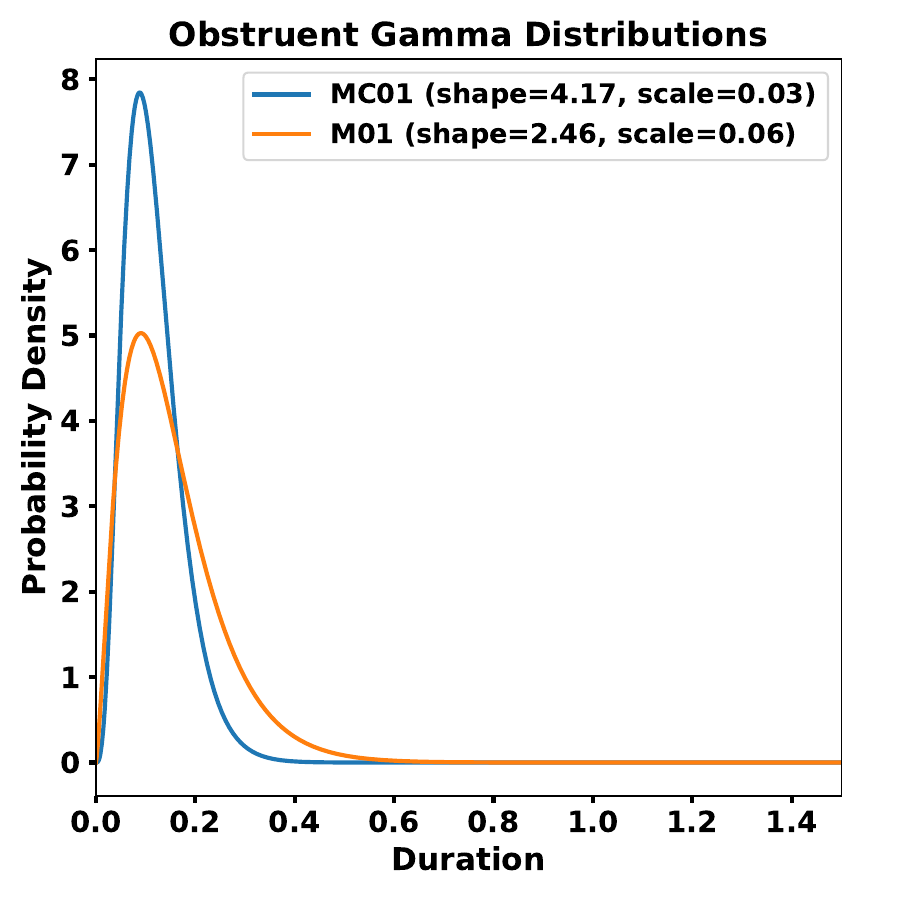}
    \caption{}
    \label{fig:obstruent_distribution}
  \end{subfigure}
  \begin{subfigure}{.49\columnwidth}
    \includegraphics[width=\columnwidth]{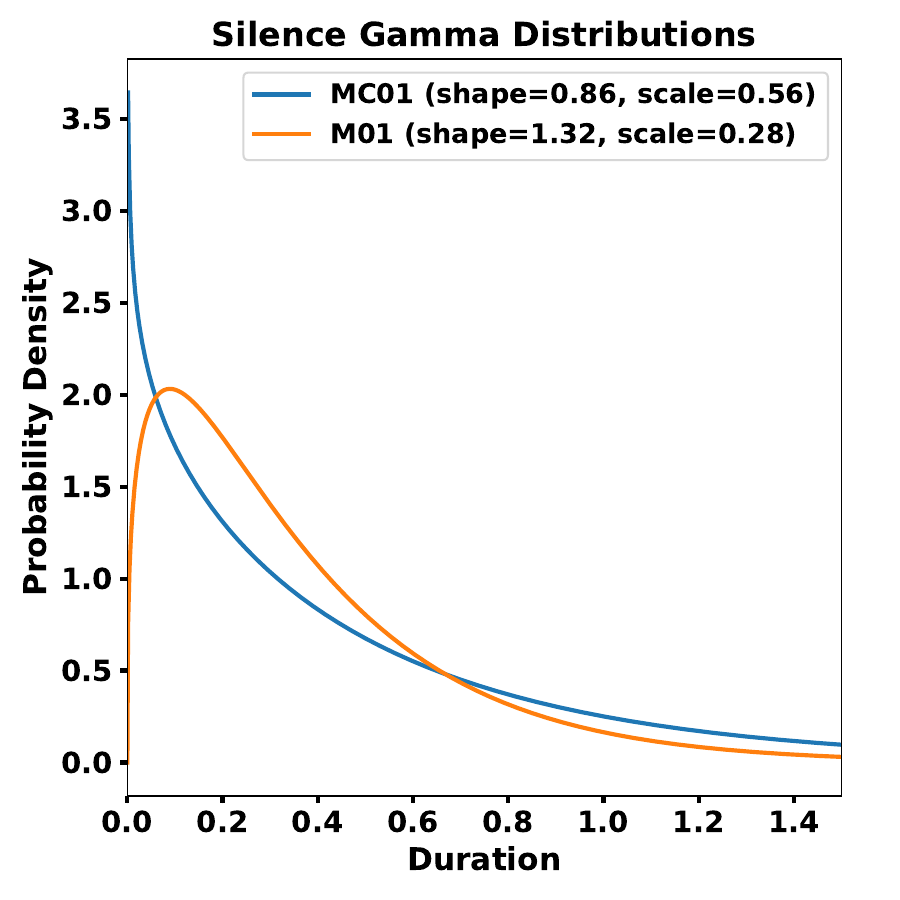}
    \caption{}
    \label{fig:silence_distribution}
  \end{subfigure}
  
  \caption{Visualization of computed rhythm models:  (a) Global speaking rates for each Torgo speaker, categorized by severity. (b-d) Comparison of gamma duration distributions per speech type for control speaker MC01 and dysarthric speaker M01.}
  \label{fig:rhythm_models}
\end{figure*}
\begin{figure}[h]
    \centering
    \includegraphics[width = .8\columnwidth]{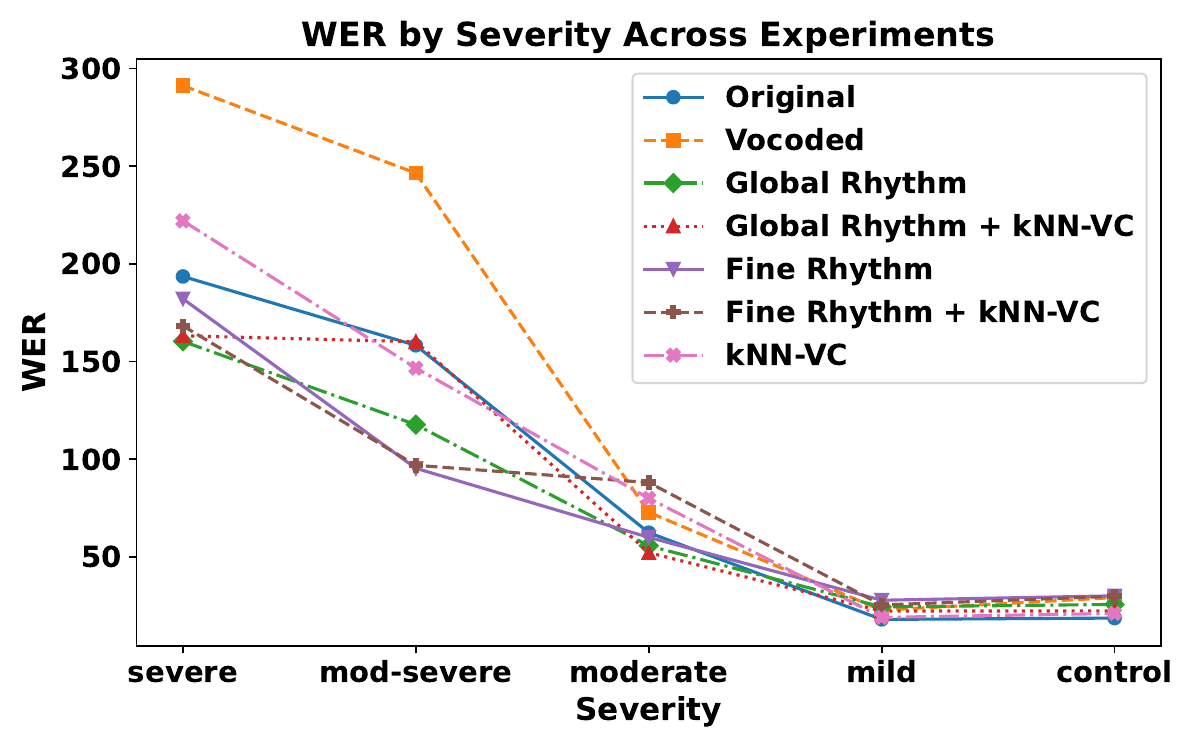}
    \caption{WER results on Torgo, grouped by severity level, presented for each experimental configuration.}
    \label{fig:wer}
\end{figure}
\vspace{-4pt}
\section{Results}

Figure \ref{fig:segmented_waveform} illustrates the segmentation obtained for a waveform from speaker M02. The segmenter successfully separates distinct parts of the waveform, accurately identifying silence and noise segments while classifying speech regions as obstruents or sonorants with reasonable, though imperfect, accuracy when compared to ground truth phonemic transcriptions. Speaking rates, shown in Figure \ref{fig:speaking_rates}, reflect expected trends across severity levels, with severe and moderately severe cases exhibiting lower rates compared to the higher rates seen in mild and control speakers. These results suggest that the segmenter's classification of sonorants serves as an effective proxy for counting syllables when estimating speaking rates in dysarthric speech. Further analysis of fine-grained rhythm distributions, depicted in Figures 3b-d, reveals that the durations of sonorants and silences differ notably between control speaker MC01 and M01, a speaker with dysarthria, with dysarthric speech tending towards longer durations. This is consistent with previous findings which show that
sonorant durations \cite{Bhat2016} and between-word pauses and silences increase with the severity of the disorder \cite{Soleymanpour2022}.

Figure \ref{fig:wer} illustrates the ASR results, with WER increasing substantially with severity, largely due to hallucinations in the Whisper model, a known issue with data outside its domain \cite{whisper_hallucinations}. Table~\ref{tab:hallucination} shows an example of hallucinated output. Performance on vocoded data is markedly worse than on original recordings, likely due to artifacts introduced by the vocoder which is not trained on dysarthric speech. kNN-VC mitigates this issue by aligning features with the vocoder’s training distribution, achieving results close to original recordings. Rhythm conversion improves outcomes further, particularly when compared to vocoded data, and global and fine-grained rhythm modeling demonstrates similar efficacy. Notably, combining rhythm conversion with kNN-VC does not yield noticeable gains over rhythm conversion alone. Consistent with this, \cite{Prananta2022} found that when generating synthetic dysarthric speech for ASR data augmentation, trained VC models also do not significantly outperform simple time-stretching approaches. Overall, conversion methods provide greater benefits as severity increases, and while rhythm conversion enables improvements over the original data, WER remains high.

\begin{table}[ht]
\centering
\resizebox{\columnwidth}{!}{
\begin{tabular}{@{}lp{6cm}@{}}
\toprule
\textbf{Utterance} & M01 Session 2\_3 0159.wav \\ 
\midrule
\textbf{Original Transcript} & This is not a program of socialized medicine. \\ 
\textbf{Whisper Transcript} & DB, that’s a program. I just, I, I just, I just, I just, I just, I just, I just, I just, I just, I just \\ 
\bottomrule
\end{tabular}
}
\caption{Example of hallucinated output generated by Whisper for a Torgo utterance.}
\label{tab:hallucination}
\end{table}

\section{Discussion and conclusions}

The segmentation results demonstrate that while our proposed method effectively identifies key waveform regions, its performance on dysarthric speech still has room for improvement, suggesting that adaptations tailored to dysarthric speech could improve accuracy. Speaking rate and rhythm distribution analyses indicate that the unsupervised methods align with known patterns of dysarthria, such as reduced speaking rates with increased severity, indicating that rhythm modeling is reasonable and the method potentially valuable beyond conversion for applications like dysarthria detection and analysis. 
ASR performance results reinforce the importance of rhythm modeling, as rhythm conversion consistently reduces WER in more severe cases, particularly by mitigating hallucinations through normalized speaking rates. Although voice conversion alone brings dysarthric speech closer to healthy domains, it does not outperform original recordings, which further highlights the importance of rhythmic conversion. However, the ASR results, while improved, still fall short of achieving low WER in more severe cases. This underscores the need for further research in converting dysarthric speech to healthy speech to enable practical usage of ASR models trained on healthy speech.
Overall, the study highlights the promise of unsupervised rhythm and voice conversion techniques for dysarthric speech. Indeed, their low data requirements and zero-shot capabilities are well suited for this domain, motivating further research in this direction. Future work should focus on domain-specific adaptations, such as improving the segmentation technique for dysarthric speech, which could enable broader use cases and benefits including rhythm modeling for dysarthric speech analysis, detection, and treatment.

\bibliographystyle{IEEEbib}
\bibliography{refs}

\end{document}